# Higher-Order Flatband-Touching Points Mediate Unconventional Bulk–Boundary Correspondence

Limin Song[1†], Jianqi Xiong[1†], Liqin Tang[1,*], Daohong Song[1,2], and Zhigang Chen[1,2,*]

[1] *The MOE Key Laboratory of Weak-Light Nonlinear Photonics, TEDA Applied Physics Institute and School of Physics, Nankai University, Tianjin 300457, China*

[2] *Collaborative Innovation Center of Extreme Optics, Shanxi University, Taiyuan, Shanxi 030006, China*

[†] *These authors contributed equally to this work*

* Corresponding author: tanya@nankai.edu.cn; zgchen@nankai.edu.cn

The bulk-boundary correspondence establishes a fundamental connection between bulk band topology and the emergence of boundary states, underpinning topological phenomena across electronic, acoustic, photonic, and many other wave systems. A broad class of singular flatband systems, however, exhibits an unconventional form of bulk-boundary correspondence, in which robust boundary modes (RBMs) originate from flatband-touching points (FBTPs) and the real-space topology associated with incomplete sets of compact localized states (CLSs), rather than conventional momentum-space bulk topological invariants. To date, experimental studies of such boundary modes have been limited to systems hosting lower-order FBTPs. Here, we experimentally demonstrate unconventional bulk-boundary correspondence in a photonic cross-bedded Kagome-like lattice featuring a higher-order FBTP formed by the touching of two degenerate flatbands and a dispersive band. Using continuous-wave laser writing, we observe shape-preserving propagation of both bulk CLSs and boundary RBMs. Remarkably, unlike previously reported flatband systems, the RBMs retain their boundary localization under input-field perturbations involving both singular and nonsingular CLSs, revealing an unconventional boundary response associated with the higher-order FBTP in this lattice. Our results establish the Kagome-like lattice as a new platform for exploring flatband physics and unconventional bulk-boundary correspondence, with potential applications in wave localization, flatband lasers, and other emerging topological photonic devices.

## 1. Introduction

Band touching is a universal phenomenon in band theory and occurs across a wide range of physical systems, including condensed matter, acoustics, and photonics. A representative example is the well-known Dirac point, which arises from the linear band touching of two dispersive bands in Dirac materials [1], such as graphene [2] and its photonic counterpart, photonic graphene [3-8]. Research has further extended the concept to tilted type-II and type-III Dirac points [9-12]. Band-touching points host a wealth of intriguing physical phenomena. Notable examples include flatband superconductivity associated with Dirac points [13], universal momentum-to-real-space topological singularity mapping [14-17], anomalous Landau level spreading [18], quantum-distance-induced interface states [19], as well as symmetric and asymmetric conical diffraction [15,20,21]. Furthermore, the closing and reopening of band gaps at band-touching points serve as hallmarks of topological phase transitions, underpinning the emergence of topological boundary states through the bulk–boundary correspondence [22,23]. Recent experimental realizations of non-Abelian Thouless pumping has also uncovered the essential role of band-touching degeneracy, where chiral-symmetry-protected states and degenerate subspaces are crucial for realizing non-Abelian transport [24,25].

In particular, flatband systems feature a unique band-touching scenario in which one of the two intersecting bands remains entirely flat across the entire Brillouin zone, as exemplified by the band structures of Kagome [26,27] and decorated honeycomb lattices [28]. In these systems, the flat and dispersive bands meet through quadratic [29,30], rather than linear, band touching, giving rise to second-order flatband-touching points (FBTPs). The FBTP refers to singularity at which all the components of flatband Bloch wavefunction vanish. In general, the order $N$ of the FBTP is determined by the number of eigenvalues that degenerates at the singularity, analogous to but conceptually distinct from the order of exceptional points in non-Hermitian systems [31-33]. $N > 2$ is generally termed higher-order FBTP, which can be formed when a single flatband intersects Dirac cones, as reported in several lattice geometries, including Lieb [34-36], super-honeycomb [37-39], fractal-like [40-42], and chiral borophene lattices [43]. These higher-order FBTPs provide a versatile platform for exploring pseudospin-1 and pseudospin-2 physics [15,43-45]. Meanwhile, non-Hermitian systems can also support symmetry-protected higher-order FBTPs [33,46]. An intriguing flatband skin effect has been uncovered recently [47,48], related to non-Hermitian higher-order FBTPs (exceptional points). In contrast, we focus on Hermitian higher-order FBTPs formed by degenerate flatbands and dispersive band and study an unconventional boundary localization.

Localized boundary states associated with FBTPs provide a promising route to uncover unconventional forms of bulk–boundary correspondence [19,29,49,50]. Theoretical studies have shown that robust boundary modes (RBMs) emerging in singular flatband systems define a distinct type of bulk–boundary correspondence, linking FBTPs in infinite bulk systems to boundary modes in their finite-sized counterparts [29,49,50]. This unconventional bulk–boundary correspondence differs fundamentally from its conventional counterpart, as evidenced by several distinctive properties of RBMs. First, unlike conventional topological boundary states that reside within bulk bandgaps, RBMs appear at the same energy as the flatband. This unusual property arises because RBMs are formed through the coherent superposition of macroscopically many bulk compact localized states (CLSs) [29,51,52]. Second, RBMs are protected by the intrinsic flatband-touching singularity and the perfect flatness of the band, rather than by bulk topological invariants. The singularity allows bulk CLSs to interfere destructively under periodic boundary conditions while forming stable RBMs under open boundaries. Meanwhile, the nondispersive flatband ensures that RBMs remain degenerate with the bulk CLSs, making them unusual bound states embedded in the flatband continuum [34]. Third, RBMs exhibit a linear dependence on bulk CLSs, a characteristic absent in conventional topological boundary modes. This linear dependence renders RBMs robustness against truncation or destruction by additional CLSs [29], in sharp contrast to conventional fragile boundary modes. Figure 1 schematically compares the different types of boundary modes and their corresponding FBTPs; representative examples can be found in Refs. [28,29,53-55]. Finally, unlike conventional topological boundary modes that feature exponential decay of localization strength, RBMs display ultra-strong localization at the lattice boundary. These distinctive properties make RBMs a subject of fundamental research interest. Whether increasing the order of the flatband-touching singularity fundamentally modifies the formation mechanism, localization, or robustness of RBMs—and hence the underlying bulk–boundary correspondence—remains an open question. Although RBMs have been observed experimentally in photonic Kagome and decorated honeycomb lattices [28,50,53], both of which host only lowest-order FBTPs, the experimental observation and systematic characterization of RBMs associated with higher-order FBTPs remain lacking (highlighted by the dashed rectangle in Fig. 1).

Here, we introduce a flatband lattice hosting a higher-order FBTP and experimentally demonstrate the associated RBMs and unconventional bulk–boundary correspondence in a photonic platform. Beyond the conventional characteristics of RBMs in lowest-order FBTP systems - namely, robustness against singular CLSs but vulnerability to nonsingular CLSs - we show that increasing the order of the flatband-touching singularity fundamentally modifies the robustness mechanism underlying unconventional bulk–boundary correspondence. Specifically, the higher-order FBTP hosts two degenerate flatbands touching a dispersive band, enabling structural compatibility between singular and nonsingular CLSs. As a consequence, the RBMs are immune to perturbations from both singular and nonsingular CLSs, in striking contrast to all previously reported flatband systems. To the best of our knowledge, neither a photonic platform hosting such a higher-order FBTP nor the associated robustness mechanism of unconventional RBM has previously been predicted or experimentally demonstrated. Our results advance the understanding of unconventional bulk–boundary correspondence in flatband systems and establish higher-order FBTPs as a versatile platform for exploring intriguing flatband physics.

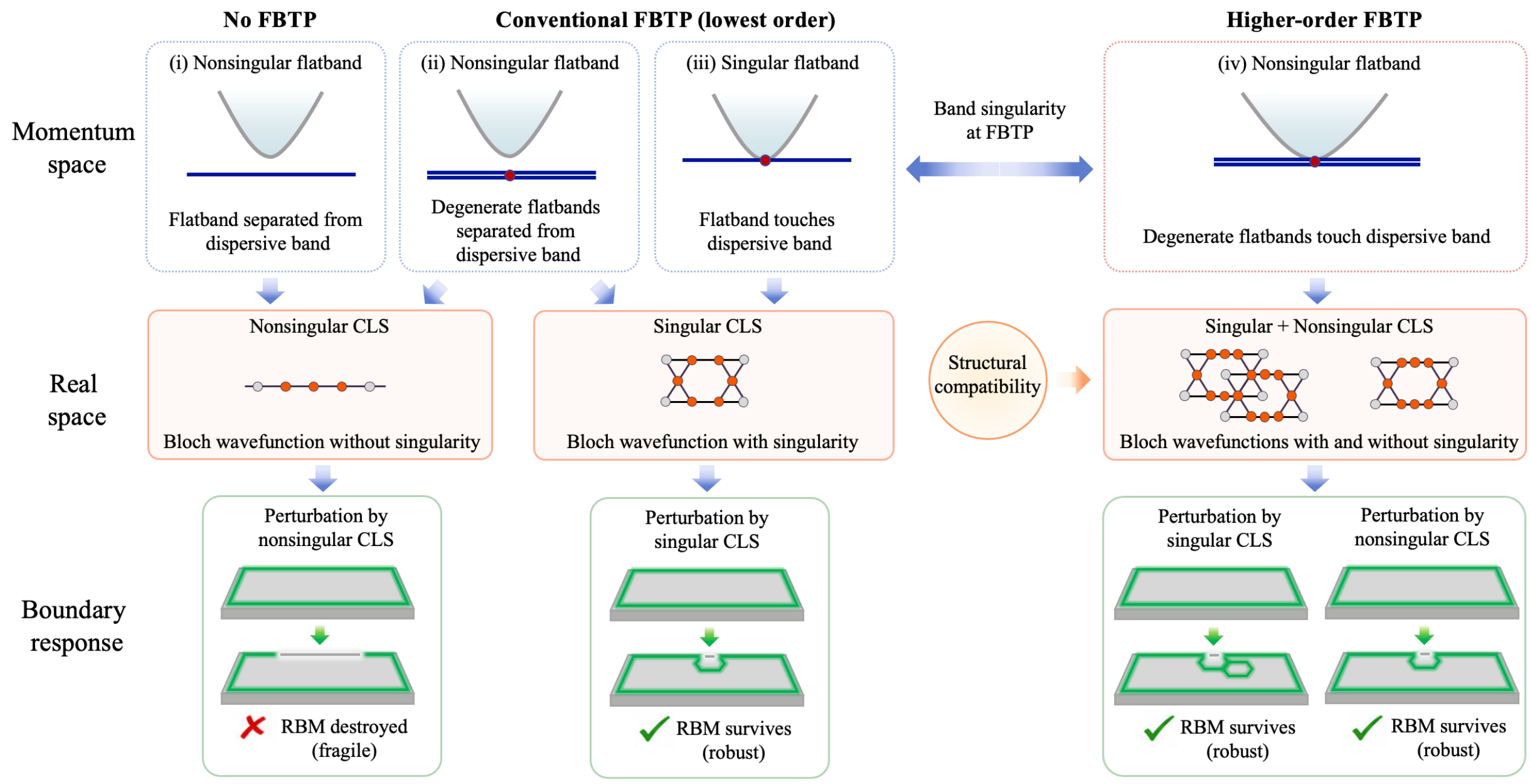

**Fig. 1. Schematic comparison of conventional and higher-order FBTP-mediated boundary modes.** The identification of robust and fragile boundary modes of generic flatband lattices (gray rectangular cuboids) is based on whether flatband boundary modes (green lines) survive perturbations induced by different compact localized states (CLSs, orange sites). For simplicity, CLSs with Bloch wavefunctions that do or do not contain momentum-space singularities are referred to as singular and nonsingular CLSs, respectively. The Bloch-wavefunction singularity originates from an FBTP (red circles) in the momentum-space band structure. (i) Spectrally isolated single flatband (nonsingular flatband, no FBTP, without RBMs). (ii) Spectrally isolated degenerate flatbands (nonsingular flatbands, second-order FBTPs, RBMs depend on CLS structure). (iii) Conventional single flatband–dispersive band touching (singular flatband, second-order FBTPs, with RBMs). (iv) Two degenerate flatbands simultaneously touching one dispersive band (nonsingular flatbands, third-order FBTPs, RBMs survive distinct CLS-type perturbations). Higher-order FBTP systems enable compatibility between singular and nonsingular CLSs, thereby protecting boundary modes against perturbations from both types of CLSs.

## 2. Results

### 2.1. Lattice and Higher-Order FBTP

To realize unconventional bulk–boundary correspondence mediated by a higher-order FBTP, we construct a bilayer-dislocated Kagome-like lattice, specifically designed to host a higher-order flatband-touching singularity. This lattice can be viewed as two Kagome-lattice ribbons laterally shifted with respect to one another, composed entirely of triangular and hexagonal motifs - the fundamental structural units of a conventional Kagome lattice [26]. However, unlike a conventional Kagome lattice, the present structure is constructed from two laterally displaced decorated Kagome ribbons, where an additional site is inserted at the midpoint of the upper and lower edges of each ribbon, as illustrated in Fig. 2(a). The relative displacement between the two decorated Kagome layers gives rise to a dislocated bilayer geometry composed of interconnected triangular and hexagonal units. Owing to this distinctive architecture, we refer to this lattice as cross-bedded Kagome-like lattice. The primitive unit cell of the lattice contains eight inequivalent sites, labeled A–H within the shaded rhombus in Fig. 2(b). Unlike twisted bilayer Kagome lattices [56,57] or AA/AB-stacked honeycomb bilayers, the Kagome-like lattice designed here does not originate from rotational stacking or conventional bilayer registry. Instead, it is formed by two dislocated decorated Kagome ribbons whose relative registry generates the characteristic lattice geometry. A photonic implementation of the Kagome-like lattice, realized using the continuous-wave laser-writing technique [36,58-60] through point-to-point optical induction in a 10-mm-long nonlinear strontium-barium-niobate (SBN) crystal, is shown in Fig. 2(c). Each site (waveguide) of the finite-size lattice structure is precisely controlled by translating the position of the writing beam via a spatial light modulator (Holoeye Pluto, 1920 × 1080 resolution). By optimizing the writing laser power (about 70 mW) and the applied electric field (1.4 kV/cm), a relatively uniform refractive-index structure defining the photonic lattice is achieved, with $\Delta n(x,y) \sim 10^{-4}$. This bilayer-dislocated geometry modifies the coupling network and gives rise to a higher-order flatband-touching singularity that is absent in conventional Kagome lattices.

The propagation of light in the waveguide array is governed by the paraxial Schrödinger-like equation [14,61]

$$i\frac{\partial \psi(x,y,z)}{\partial z} = -\left(\frac{1}{2k_0 n_0}\nabla_\perp^2 + k_0 \Delta n(x,y)\right)\psi(x,y,z) = H_{\text{cont}}\psi(x,y,z), \quad (1)$$

where $z$ is the longitudinal propagation distance, $\psi(x,y,z)$ is the optical field envelope, $k_0 = 2\pi n_0/\lambda$ is the wavenumber in the medium, $\lambda = 532$ nm is the wavelength both for laser-writing and probing, $n_0 = 2.35$ is the background refractive index of the material in which the lattice is embedded, and $\nabla_\perp^2 = \partial_x^2 + \partial_y^2$ is the paraxial-diffraction operator. Here $H_{\text{cont}}$ denotes the continuum Hamiltonian governing wave propagation in the photonic lattice.

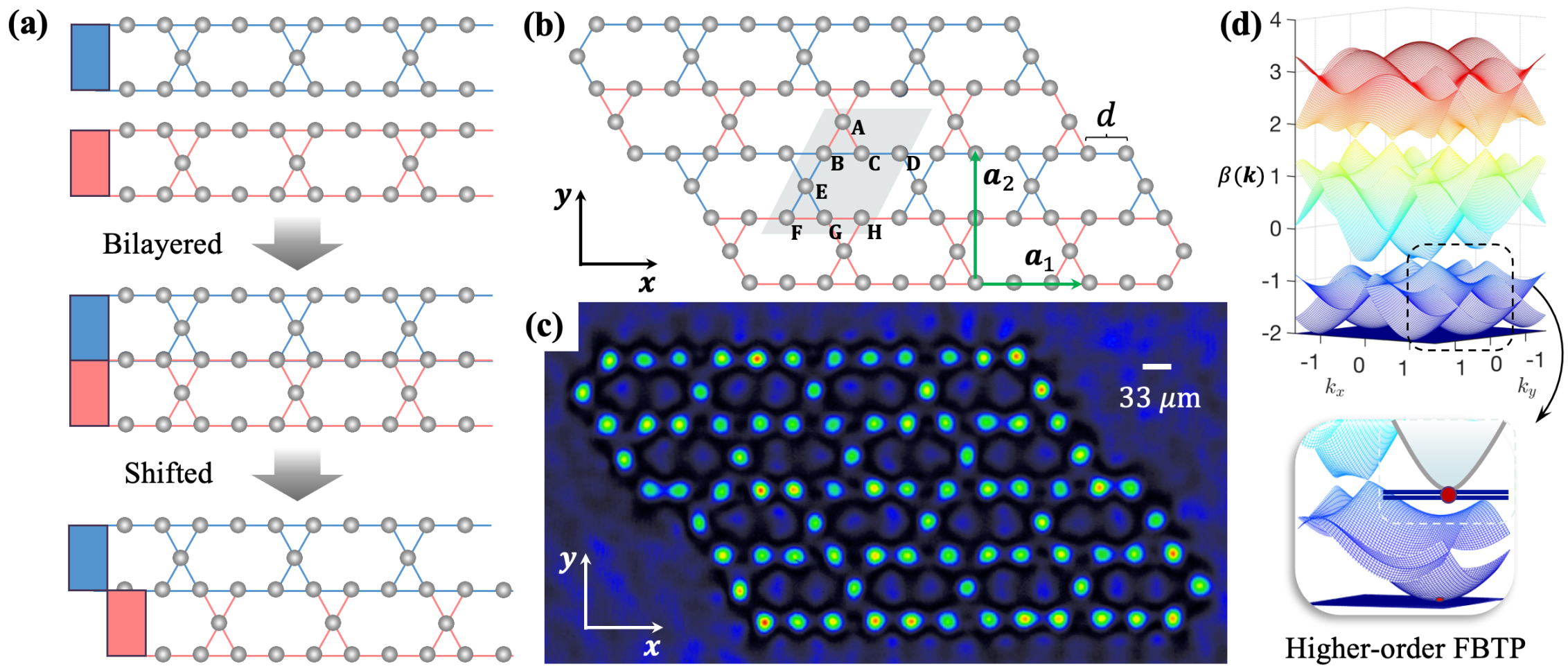

**Fig. 2. Lattice construction and band structure featuring a higher-order FBTP.** (a) Schematic showing the construction of the cross-bedded Kagome-like lattice from two laterally shifted decorated Kagome ribbons. (b) Primitive unit cell of the lattice containing eight inequivalent sublattice sites (A–H). Green arrows denote the primitive lattice vectors $\boldsymbol{a}_1 = a(3,0)$ and $\boldsymbol{a}_2 = a(0,2\sqrt{3})$. $a = 1$ for theoretical analysis, corresponding to 99 μm in real lattice. $d$ is the spacing between nearest-neighbor sites, which defines the nearest-neighbor hopping amplitude $t$ (normalized to $t = 1$). (c) Experimental laser-written photonic implementation of the lattice with a nearest-neighbor site spacing of approximately 33 μm. (d) Tight-binding band structure of the infinite Kagome-like lattice, consisting of six dispersive bands and two degenerate flatbands. Bottom: Enlarged view of the higher-order FBTP, which mediates the unconventional bulk–boundary correspondence explored in this work.

Within the tight-binding approximation, evanescent coupling between waveguides can be described by a discrete lattice model [61,62]. When considering only the nearest-neighbor coupling, the Bloch Hamiltonian of the system takes the form

$$H = \begin{pmatrix} M_d & M_{od} \\ M_{od}^{\mathrm{T}} & M_d \end{pmatrix}, \tag{2}$$

where $M_d$ and $M_{od}$ denote the diagonal and off-diagonal submatrices, respectively, with their specific forms given as follows

$$M_d = t\begin{pmatrix} 0 & 1 & 1 & 0 \\ 1 & 0 & 1 & e^{-i\boldsymbol{k}\boldsymbol{a}_1} \\ 1 & 1 & 0 & 1 \\ 0 & e^{i\boldsymbol{k}\boldsymbol{a}_1} & 1 & 0 \end{pmatrix}, \tag{3}$$

$$M_{od} = t\begin{pmatrix} 0 & 0 & e^{i\boldsymbol{k}\boldsymbol{a}_2} & e^{i\boldsymbol{k}\boldsymbol{a}_2} \\ 1 & 0 & 0 & 0 \\ 0 & 0 & 0 & 1 \\ e^{i\boldsymbol{k}\boldsymbol{a}_1} & 0 & 0 & 0 \end{pmatrix}. \tag{4}$$

The bilayer-dislocated geometry of the cross-bedded Kagome-like lattice is fully captured by the block Hamiltonian $H$. In Eqs. (3) and (4), $t$ represents the nearest-neighbor hopping amplitude, and $\boldsymbol{k} =$ $(k_x, k_y)$ is the transverse Bloch wave vector.

Diagonalizing the Hamiltonian $H$ yields the full energy spectrum of the cross-bedded Kagome-like lattice system, which comprises eight energy bands, as illustrated in Fig. 2(d). Specifically, the spectrum contains two doubly degenerate flatbands at $\beta(\boldsymbol{k}) = -2$ and six dispersive bands. Notably, these flatbands touch the adjacent dispersive band quadratically at a common high-symmetry point (red circle) in the first Brillouin zone, giving rise to a higher-order FBTP. This band singularity distinctly differentiates the cross-bedded Kagome-like lattice from conventional two-dimensional flatband lattices, including Kagome [26,27,53], Lieb [34-36], super-honeycomb [38,39], and fractal-like lattices [40-42,54]. In the conventional Kagome lattice, a single flatband touches a dispersive band to form a quadratic band touching [case (iii) in Fig. 1]. By contrast, Lieb and super-honeycomb lattices host a single singular flatband intersecting dispersive bands, giving rise to pseudospin-1 conical band crossings. In all three lattice systems, the unconventional bulk–boundary correspondence is associated with a single singular flatband originating from an immovable Bloch-wavefunction singularity [29,49]. More recently, fractal-like lattices are theoretically and experimentally shown to host degenerate flatbands that include both singular and nonsingular flatbands [41]. However, the degenerate flatbands in fractal-like lattices are spectrally isolated [case (ii) in Fig. 1], and their RBMs depend critically on the choice of CLS set spanning the flatband subspace [54]. By contrast, the higher-order FBTP of the cross-bedded Kagome-like lattice simultaneously couples two degenerate flatbands to a dispersive band, fundamentally modifying the unconventional bulk–boundary correspondence. As a consequence, the RBMs exist regardless of the choice of CLS set, including both singular and nonsingular CLS configurations associated with the degenerate flatbands [case (iv) in Fig. 1].

### 2.2. Singular and Nonsingular CLSs

We first investigate the two distinct types of CLSs emerging in the proposed lattice system. By solving the eigenvalue equation, $H|v(\boldsymbol{k})\rangle = \beta|v(\boldsymbol{k})\rangle$, we derive the normalized Bloch wavefunctions corresponding to the doubly degenerate flatbands $\beta = -2$,

$$|v_1(\boldsymbol{k})\rangle = \frac{1}{\alpha_{\boldsymbol{k}}^{(1)}} \begin{pmatrix} -e^{i\boldsymbol{k}\boldsymbol{a}_2} - e^{i\boldsymbol{k}(\boldsymbol{a}_1+\boldsymbol{a}_2)} \\ -1 + e^{i\boldsymbol{k}\boldsymbol{a}_2} \\ 1 + e^{i\boldsymbol{k}(\boldsymbol{a}_1+\boldsymbol{a}_2)} \\ -1 - e^{i\boldsymbol{k}(\boldsymbol{a}_1+\boldsymbol{a}_2)} \\ e^{-i\boldsymbol{k}\boldsymbol{a}_1} + 1 \\ -1 - e^{-i\boldsymbol{k}\boldsymbol{a}_1} \\ 0 \\ 1 + e^{i\boldsymbol{k}\boldsymbol{a}_1} \end{pmatrix}, \tag{5}$$

$$|v_2(\boldsymbol{k})\rangle = \frac{1}{\alpha_{\boldsymbol{k}}^{(2)}} \begin{pmatrix} -e^{ik\boldsymbol{a}_2} - e^{ik(\boldsymbol{a}_1+\boldsymbol{a}_2)} \\ e^{ik\boldsymbol{a}_1} + e^{ik\boldsymbol{a}_2} \\ -e^{ik\boldsymbol{a}_1} + e^{ik(\boldsymbol{a}_1+\boldsymbol{a}_2)} \\ e^{ik\boldsymbol{a}_1} - e^{ik(\boldsymbol{a}_1+\boldsymbol{a}_2)} \\ -1 - e^{ik\boldsymbol{a}_1} \\ 0 \\ 1 + e^{ik\boldsymbol{a}_1} \\ 0 \end{pmatrix}, \tag{6}$$

where $\alpha_{\boldsymbol{k}}^{(1)} = \sqrt{2}\sqrt{7 + 4\cos(\boldsymbol{k}\cdot\boldsymbol{a}_1) - \cos(\boldsymbol{k}\cdot\boldsymbol{a}_2) + 2\cos(\boldsymbol{k}\cdot\boldsymbol{a}_1 + \boldsymbol{k}\cdot\boldsymbol{a}_2)}$ and $\alpha_{\boldsymbol{k}}^{(2)} = \sqrt{2}\sqrt{6 + 3\cos(\boldsymbol{k}\cdot\boldsymbol{a}_1) - 2\cos(\boldsymbol{k}\cdot\boldsymbol{a}_2) + \cos(\boldsymbol{k}\cdot\boldsymbol{a}_1 - \boldsymbol{k}\cdot\boldsymbol{a}_2)}$ are the normalization factors [29,49]. These Bloch wavefunctions enable the construction of 14-site and 12-site CLSs extending over four unit cells (two hexagonal plaquettes) in real space, forming localized states embedded within the bulk continuum. Two representative CLS configurations are identified: a step-shaped state associated with $|v_1(\boldsymbol{k})\rangle$ (CLS-1) and a rhombic state associated with $|v_2(\boldsymbol{k})\rangle$ (CLS-2), as shown in Figs. 3(a1) and 3(b1), respectively. The red and blue lattice sites possess identical wavefunction amplitudes but carry opposite phases of 0 and π. Specifically, CLS-1 has zero amplitude on all lattice sites belonging to the G sublattice, whereas CLS-2 has vanishing amplitude on all lattice sites of the F and H sublattices, as given by Eqs. (5) and (6). These distinct sublattice occupation patterns are directly encoded in the analytical Bloch wavefunctions and their corresponding real-space CLS profiles. Furthermore, the Bloch wavefunctions of CLS-1 and CLS-2 share the same singular momentum point at $\boldsymbol{k} = (\pi, 0)$. This inherent singularity leads to an incomplete basis representation of the flatband subspace, resulting missing eigenstates when adopting CLS-1 and CLS-2 as the flatband basis set. The singular nature of these Bloch wavefunctions is quantitatively manifested in their momentum-dependent normalization factors [29,49], as illustrated in Figs. 3(c) and 3(d), which demonstrate identical singular behaviors for $|v_1(\boldsymbol{k})\rangle$ and $|v_2(\boldsymbol{k})\rangle$. As discussed above, these FBTP-induced singularities constitute the fundamental origin of the RBMs in the cross-bedded Kagome-like lattice.

The lattice also supports two additional hexagonal CLSs, termed CLS-3 and CLS-4, as depicted in Figs. 3(a2) and 3(a3), respectively. The corresponding Bloch wavefunctions of these two hexagonal CLSs can be explicitly formulated as

$$|v_3(\boldsymbol{k})\rangle = \frac{1}{\alpha_{\boldsymbol{k}}^{(3)}} \begin{pmatrix} -e^{ik\boldsymbol{a}_2} - e^{ik(\boldsymbol{a}_1+\boldsymbol{a}_2)} \\ e^{ik\boldsymbol{a}_2} \\ e^{ik(\boldsymbol{a}_1+\boldsymbol{a}_2)} \\ -e^{ik(\boldsymbol{a}_1+\boldsymbol{a}_2)} \\ 0 \\ -1 \\ 1 \\ e^{ik\boldsymbol{a}_1} \end{pmatrix}, \tag{7}$$

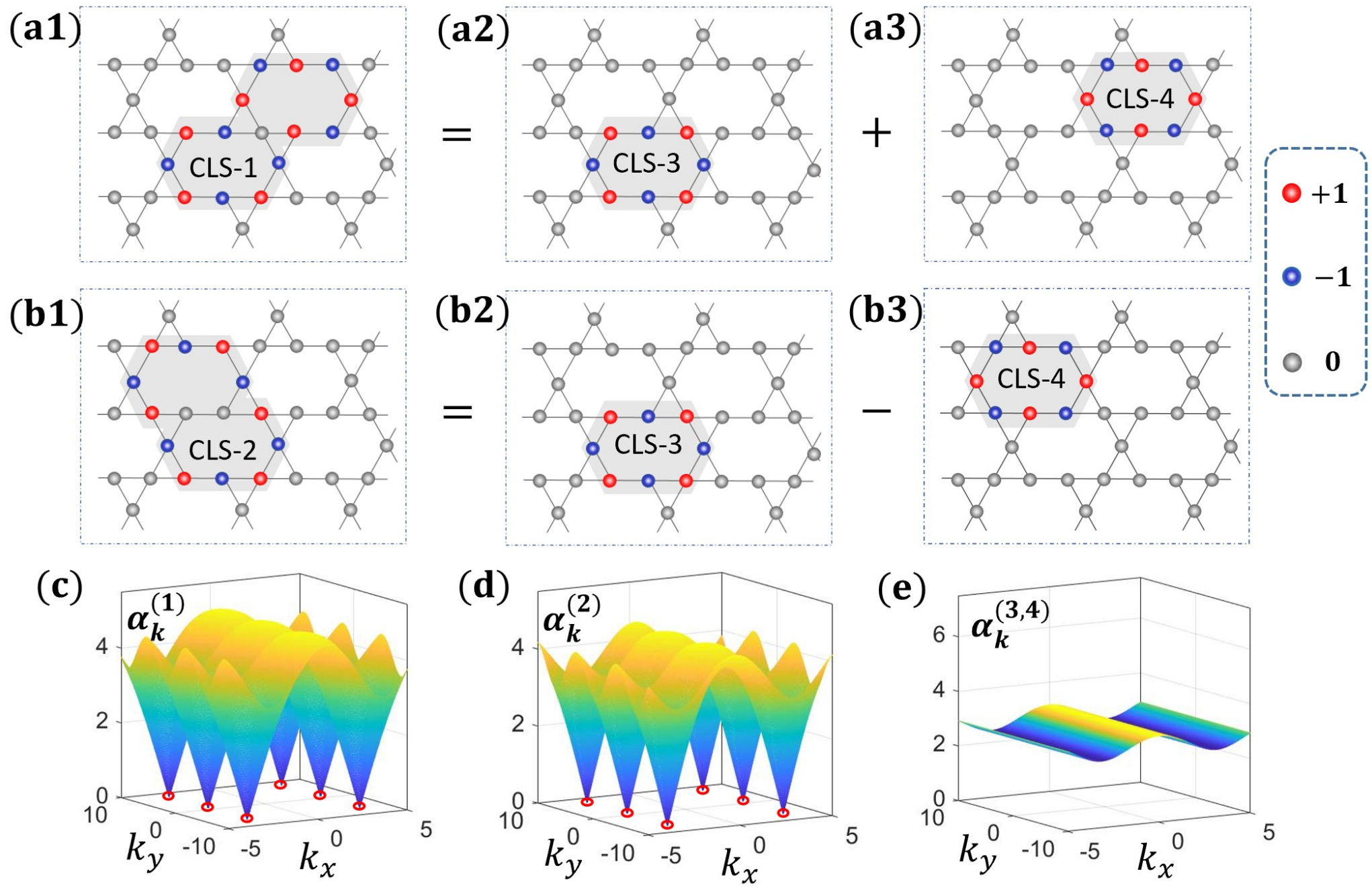

**Fig. 3. Theoretical analysis of the CLSs hosted by the degenerate flatbands associated with higher-order FBTP and their structural compatibility.** (a1, b1) Two singular CLSs with step-shaped and rhombic profiles, corresponding to CLS-1 (described by $|v_1(\boldsymbol{k})\rangle$) and CLS-2 (described by $|v_2(\boldsymbol{k})\rangle$), respectively. Remarkably, each CLS can be decomposed into two smaller hexagonal CLSs, namely CLS-3 ($|v_3(\boldsymbol{k})\rangle$) and CLS-4 ($|v_4(\boldsymbol{k})\rangle$). Accordingly, CLS-1 and CLS-2 can be explicitly expressed as linear superpositions of CLS-3 and CLS-4, as demonstrated in (a2, a3) and (b2, b3). Although CLS-3 and CLS-4 differ only subtly in geometry, they can be distinguished by the arrangement of the surrounding triangular plaquettes. The red and blue dots denote lattice sites with normalized amplitudes of +1 and −1, respectively, while the gray dots indicate zero amplitude. (c)-(e) Momentum-space distributions of the normalization factors $\alpha_k^{(i)}$ $(i = 1,2,3,4)$ as functions of $k_x$ and $k_y$. The zeros of $\alpha_k^{(1,2)}$ indicate the discontinuities of the corresponding singular Bloch wavefunctions $|v_1(\boldsymbol{k})\rangle$ and $|v_2(\boldsymbol{k})\rangle$, giving rise to incomplete flatband spanning sets. By contrast, $\alpha_k^{(3,4)}$ remain finite throughout the Brillouin zone, indicating nonsingular Bloch wavefunctions $|v_3(\boldsymbol{k})\rangle$ and $|v_4(\boldsymbol{k})\rangle$, which form complete flatband spanning sets.

$$|v_4(\boldsymbol{k})\rangle = \frac{1}{\alpha_{\boldsymbol{k}}^{(4)}}\begin{pmatrix} 0 \\ 1 \\ -1 \\ 1 \\ -1 - e^{-i\boldsymbol{k}\boldsymbol{a}_1} \\ e^{-i\boldsymbol{k}\boldsymbol{a}_1} \\ 1 \\ -1 \end{pmatrix}, \tag{8}$$

where $\alpha_{\boldsymbol{k}}^{(3)} = \alpha_{\boldsymbol{k}}^{(4)} = \sqrt{8 + 2\cos(\boldsymbol{k} \cdot \boldsymbol{a}_1)}$. As illustrated in Fig. 3(e), the coefficients $\alpha_{\boldsymbol{k}}^{(3,4)}$ remain finite (nonzero) throughout the Brillouin zone, demonstrating that CLS-3 and CLS-4 constitute a complete spanning set for the flatband subspace. Accordingly, the degenerate flatbands belong to the class of nonsingular flatbands, owing to the existence of complete CLS bases spanning the entire flatband manifold [29]. This behavior is analogous to that reported for Kagome-3 [26,29] and fractal-like lattices [41,42,54]. Consequently, the Bloch-wavefunction singularities associated with $|v_1(\boldsymbol{k})\rangle$ and $|v_2(\boldsymbol{k})\rangle$ are removable [29,54], albeit they share common locations and are intrinsically different from that of the isolated degenerate flatbands [case (ii) in Fig. 1]. Notably, CLS-1, CLS-2, and their arbitrary linear superpositions can be fully reconstructed via the linear combination of CLS-3 and CLS-4 [Figs. 3(a1-a3) and 3(b1-b3)]. In contrast, neither CLS-3 nor CLS-4 can be retrieved through any linear superposition of CLS-1 and CLS-2. This asymmetric spanning relation reflects the structural compatibility between the singular and nonsingular CLS families, a unique feature of the higher-order FBTP hosted by the cross-bedded Kagome-like lattice. As shown below, the additional band touching between degenerate flatbands and dispersive band introduces a new robustness mechanism for RBMs, fundamentally enriching the unconventional bulk–boundary correspondence in flatband systems.

### 2.3. Observation of Bulk Compact Localized States

We experimentally investigate the localized propagation of the theoretically predicted bulk-localized states, i.e., the CLSs presented in Fig. 3. For flatband mode excitations, light fields remain tightly confined to the initially excited lattice sites without spreading to neighboring sites [63-65]. Such localized propagation behaviors have been experimentally observed in various photonic lattice platforms, including Lieb [34,35], Kagome [27], fractal-like [41,42], and super-Kagome [66] lattices. However, owing to the unique higher-order FBTP of the cross-bedded Kagome-like lattice and the central role of CLSs as the elementary building blocks of RBMs, experimental verification of the predicted CLSs is essential.

We conduct two sets of experiments to characterize the two classes of CLSs supported by the cross-bedded Kagome-like lattice: the singular CLS-1 [Fig. 3(a1)] and CLS-2 [Fig. 3(b1)], and the nonsingular CLS-3 [Fig. 3(c1)] and CLS-4 [Fig. 3(d1)]. A series of Gaussian beams with a wavelength of $\lambda = 532$ nm is used for excitation, the same wavelength with the writing beam. Their quasi-nondiffractive propagation zone fully covers the entire 10 mm length of the used nonlinear crystal. These Gaussian beams are coherently superposed to construct the desired probe beam. The experimental cross-bedded Kagome-like lattice consists of a finite array of waveguides with a uniform nearest-neighbor spacing of approximately 33 μm. Assisted by a spatial light modulator, the probe beam is customized to precisely replicate the real-space intensity profiles of target CLSs, together with their characteristic alternating phase distributions among adjacent lattice sites. By injecting the structured probe beam into the photonic cross-bedded Kagome-like lattice, we achieve excitation of the desired eigenvalue at the higher-order FBTP by directly matching the corresponding eigenmode, and then directly observe the propagation dynamics of these bulk flatband states. Therefore, our experiment probes real-space mode propagation related to the higher-order FBTP instead of momentum-resolved band spectroscopy.

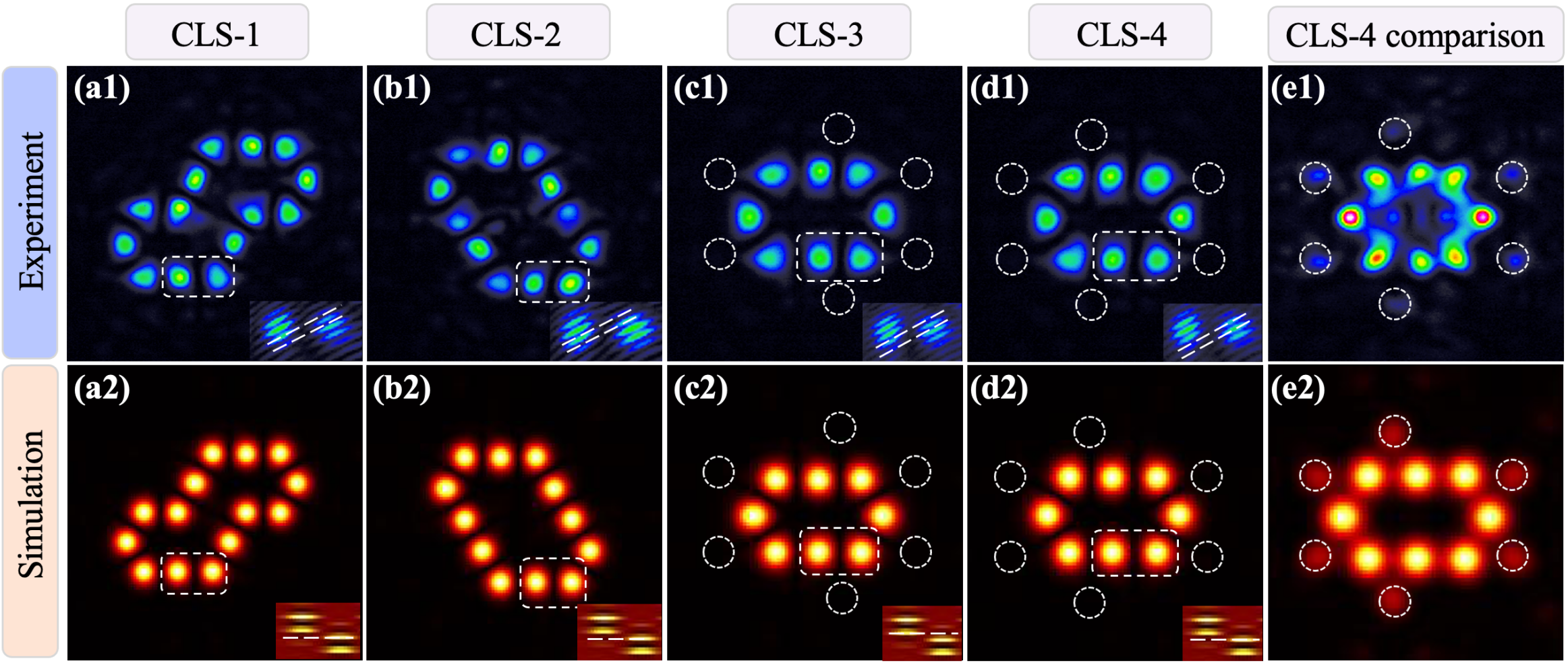

**Fig. 4. Experimental observation of bulk compact localized states (CLSs).** (a1, a2) Experimental output intensity distribution and corresponding numerical simulation for the step-shaped CLS-1. The input excitation pattern and probe-beam phase distribution are identical to those shown in Fig. 3(a1). (b1–d2) Same as (a1, a2), but for excitation of the rhombic CLS-2 [(b1, b2)], hexagonal CLS-3 [(c1, c2)], and hexagonal CLS-4 [(d1, d2)]. (e1, e2) Experimental and numerical results for an in-phase (zero relative phase difference among adjacent excited lattice sites) excitation of CLS-4, shown as a reference case without compact localization. White dashed circles indicate the neighboring lattice sites that distinguish CLS-3 from CLS-4 despite their similar geometries. Insets show interferograms from the highlighted regions, confirming the phase distributions of the corresponding CLSs.

The experimental observations are summarized in Fig. 4. In all measurements, the structured probe beam is injected into preselected mode-matched waveguides. For clarity, the input beam profiles are omitted because they closely match the corresponding output intensity distributions for individual flatband excitations. Selective excitation of CLS-1, CLS-2, CLS-3, and CLS-4 leads to output intensity profiles that remain well preserved throughout propagation, demonstrating compact localization without observable diffraction [Figs. 4(a1)–4(d1)]. The corresponding interferograms confirm the characteristic $\pi$-phase difference between neighboring excited sites. By contrast, an in-phase excitation produces pronounced discrete diffraction, with optical energy spreading into adjacent waveguides, as shown in Fig. 4(e1). We note that the two excitation schemes in Figs. 4(d1) and 4(e1) excite exactly the same set of lattice sites; however, their phase distributions determine whether the input couples to non-dispersive flatband eigenmodes (out-of-phase, compact localization) or dispersive propagating modes (in-phase, strong diffraction). These experimental results are in excellent agreement with the numerical simulations in Figs. 4(a2)–4(e2).

**2.4. Observation of Boundary Localized States: RBMs**

We now turn to the observation of RBMs and the experimental demonstration of unconventional bulk–boundary correspondence. Under periodic boundary conditions, the linear superposition of all singular CLSs yields a zero-field distribution. In contrast, for open boundary conditions, this superposition cancels throughout the bulk while leaving finite field amplitudes at the lattice boundaries, thereby generating RBMs. In our experiments, the RBM probe beam is constructed via the superposition of multiple Gaussian beams with precisely customized phase distributions, matching the theoretical RBM profile illustrated in Fig. 5(a1). After propagating through the 10-mm-long photonic lattice, the structured probe beam preserves its original spatial profile, as depicted in Fig. 5(a2). By contrast, in-phase excitation of the identical boundary configuration induces prominent discrete diffraction [61], leading to visible distortion of the output pattern and optical energy leakage into adjacent bulk lattice sites [Fig. 5(a3)]. These observations verify the strong localization of the higher-order FBTP-derived boundary states and provide direct experimental evidence for the intrinsic RBM supported by the cross-bedded Kagome-like lattice.

Furthermore, the robustness of the RBM is systematically examined under three types of boundary perturbations, including additional singular and nonsingular CLS superpositions and partial boundary excitation deficiency. When perturbed by the superposition of an additional singular CLS-2 [Fig. 5(b1)] or nonsingular CLS-4 [Fig. 5(c1)], the RBM undergoes only boundary-profile reconstruction while maintaining excellent spatial localization [Figs. 5(b2) and 5(c2)]. Even when part of the boundary excitation is intentionally removed [Fig. 5(d1)], the RBM exhibits pronounced self-healing behavior [Fig. 5(d2)], restoring a nearly perfect localized profile consistent with the unperturbed case in Fig. 5(a2). For clarity, we adopted mode fidelity as a metric to quantify the overlap of the two optical intensity profiles in experiment. For discrete lattice-site peak intensities $\{I_{1,i}\}$ [reference, Fig. 5(a2)] and $\{I_{2,i}\}$ [perturbed, Fig. 5(d2)], the intensity-based fidelity can be defined as

$$F = \frac{\left(\sum_i \sqrt{I_{1,i}}\sqrt{I_{2,i}}\right)^2}{\left(\sum_i I_{1,i}\right)\left(\sum_i I_{2,i}\right)}. \tag{9}$$

Here $I_{1,i}$ and $I_{2,i}$ are the peak intensities at the $i$-th lattice site for the reference and perturbed RBMs, respectively. Fidelity ranges from 0 to 1, with $F = 1$ representing perfect overlap between two intensity distributions. The obtained $F = 0.9975$ quantitively shows that the overall intensity profile of the perturbed boundary mode closely matches the unperturbed reference. In all three cases, in-phase excitation still leads to strong diffraction and inter-site energy coupling, as shown in Figs. 5(b3)–5(d3). All experimental measurements are well reproduced by numerical simulations, which validate the field evolution behaviors for both out-of-phase localized modes and in-phase diffusive modes, as summarized in Figs. 5(a4)–5(d4) and Figs. 5(a5)–5(d5), confirming both the localized propagation of the RBMs and the diffractive evolution of the in-phase excitations.

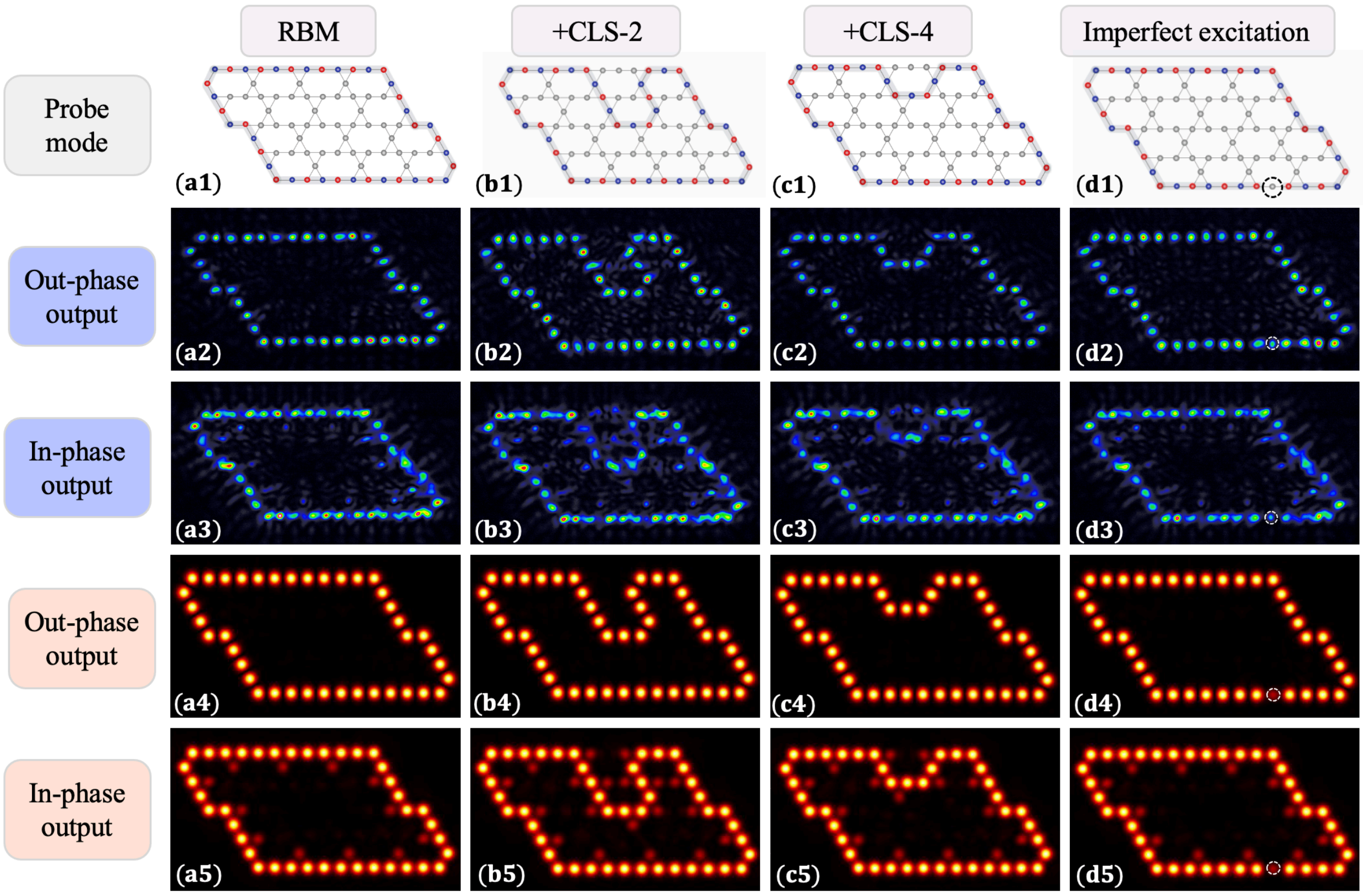

**Fig. 5. Experimental observation of robust boundary modes (RBMs)**. (a1) Amplitude and phase profile of an RBM constructed by the linear superposition of eight CLS-2 configurations, giving rise to finite boundary amplitudes. Red and blue dots denote lattice sites with normalized amplitudes of +1 and −1, while gray dots indicate zero amplitude. (a2, a3) Experimental output intensity distributions under structured out-phase (adjacent excited lattice sites carry a relative phase difference of π) and in-phase (zero relative phase difference) excitations, respectively. (a4, a5) Corresponding numerical simulations. Panels (b1–d5) follow the same layout as (a1–a5), with boundary perturbations introduced by adding CLS-2, adding CLS-4, or removing a single excitation site. The RBMs remain strongly localized under perturbations associated with both singular and nonsingular CLSs and further exhibit self-healing in the presence of local amplitude defects.

It is worth emphasizing that these perturbations act on the input excitation field via CLS superposition or partial removal of excitation sites, without modifying the underlying lattice Hamiltonian, waveguide geometry, or inter-waveguide hopping. The demonstrated robustness describes the response of flatband boundary modes to field-level CLS-type perturbations, which should be distinguished from robustness against structural imperfections or hopping-amplitude disorder in conventional topological systems [59,60,67-71]. Most importantly, unlike previously reported singular flatband systems, the RBMs retain their boundary localization under input-field perturbations involving both singular and nonsingular CLSs. These observations provide direct experimental evidence of an unconventional boundary response associated with the higher-order FBTP in the present lattice.

## 3. Discussion and Conclusion

Unlike conventional measures of localization, such as the inverse participation ratio and Lyapunov exponent, the localization of the flatband states studied here can be assessed directly by examining whether the output light remains confined to the initially excited lattice sites. To quantitatively compare the bulk/boundary confinement of the CLSs/ RBMs in observations of Figs. 4 and 5 under different excitation cases, we can calculate the localization fraction $\eta$, defined as the ratio of total optical intensity confined within the predefined lattice sites to the integrated intensity over all sampled sites,

$$\eta_{\mathrm{CLS/RBM}} = \frac{\sum_{i \in \mathrm{CLS/RBM}} I_i}{\sum_{\mathrm{all}\ i} I_i}, \tag{10}$$

where $I_i$ denotes the peak intensity at the $i$-th lattice site. A localization fraction near unity signifies that the majority of optical energy is confined within the target lattice region. For the out-of-phase excitation, the localization fraction of the CLS/RBM is close to 1, whereas the value for in-phase excitation is markedly lower.

In general, although flatband boundary modes can be synthesized via superposition of all bulk CLSs in nonsingular flatband systems, such synthesized modes differ fundamentally from genuine RBMs. For example, in the bilayer square lattice [29], these artificial boundary modes are fragile and readily destroyed upon introducing a single boundary-adjacent CLS [case (i) in Fig. 1]. By contrast, the RBM hosted by the higher-order FBTP of the cross-bedded Kagome-like lattice retains boundary localization under input-field perturbations involving both singular and nonsingular CLSs [case (iv) in Fig. 1]. This distinctive behavior is associated with the structural compatibility between the two classes of CLSs in the present higher-order-FBTP lattice. It also distinguishes the present system from fractal-like lattices [54], where RBMs associated with isolated nonsingular degenerate flatbands can still be destroyed by introducing a nonsingular CLS [case (ii) in Fig. 1]. An important point is that, although the degenerate flatbands of the cross-bedded Kagome-like lattice are not spectrally isolated, they nevertheless belong to the class of nonsingular flatbands because they admit complete nonsingular CLS spanning sets. We emphasize, however, that higher-order flatband touching does not necessarily imply nonsingular flatbands; rather, the cross-bedded Kagome-like lattice represents a specific realization in which these two features coexist. These findings provide new insight into unconventional bulk–boundary correspondence in systems featuring higher-order FBTPs.

In conclusion, we have theoretically predicted and experimentally demonstrated an unconventional bulk-boundary correspondence mediated by a higher-order FBTP in a photonic cross-bedded Kagome-like lattice. Using two-dimensional waveguide arrays fabricated by continuous-wave laser writing, we experimentally observed distinct families of bulk CLSs together with boundary RBMs associated with the higher-order FBTP. Unlike previously investigated flatband systems with lower-order FBTPs, the RBMs reported here retain their boundary localization under input-field perturbations involving both singular and nonsingular CLSs, a distinctive behavior associated with the structural compatibility of the CLS families in this lattice. The excellent agreement between theoretical analysis, numerical simulations, and experimental observations supports the proposed unconventional bulk-boundary correspondence. More broadly, our work illustrates how a higher-order flatband-touching configuration can be associated with unconventional boundary-mode behavior beyond that observed in previously studied flatband systems. Our platform also opens opportunities to investigate mode mixing under perturbations, the response of multifold degeneracy to modified lattice symmetries or couplings, and possible links to multiband wavefunction geometry. Beyond introducing a new flatband platform, these results deepen the understanding of singular and nonsingular flatband physics and open new opportunities for exploring unconventional wave localization and topological phenomena in photonics, acoustics, ultracold atoms, and other synthetic wave systems.

**Conflict of interest.** The authors declare no conflicts of interest.

**Acknowledgments.** This research was supported by the National Key R&D Program of China (No. 2022YFA1404800); the National Natural Science Foundation of China (No. W2541003, 12504384, 12134006, 12374309, and 12274242); the Natural Science Foundation of Tianjin for Distinguished Young Scientists (No. 21JCJQJC00050); the Postdoctoral Fellowship Program of CPSF (No. GZC20252237); and the China Postdoctoral Science Foundation (No. 2025M773391).

**Author contributions.** L.S. and J.X. performed the simulation and experiments. L.T., D.S., and Z.C. discussed the results. Z.C. and L.T. supervised the project. All authors contributed to this work.

**Data availability.** All data that support the findings of this study are included within the article.